\documentstyle[12pt]{article}
\catcode`\@=11

\@addtoreset{equation}{section}
\def\theequation{\arabic{section}.\arabic{equation}}

\catcode`\@=11

\def\section{\@startsection{section}{1}{\z@}{3.5ex plus 1ex minus
   .2ex}{2.3ex plus .2ex}{\large\bf}}

\def\eqnarray{\let\@currentlabel=\theequation\refstepcounter{equation}
    \global\@eqnswtrue
    \global\@eqcnt\z@\tabskip\@centering\let\\=\@eqncr
    $$\halign to \displaywidth\bgroup\@eqnsel\hskip\@centering
      $\displaystyle\tabskip\z@{##}$&\global\@eqcnt\@ne
       \hfil${{}##{}}$\hfil
      &\global\@eqcnt\tw@ $\displaystyle\tabskip\z@{##}$\hfil
       \tabskip\@centering&\llap{##}\tabskip\z@\cr}
\def\lefteqn#1{\hbox to 4\arraycolsep{$\displaystyle #1$\hss}}
\def\thesection{\arabic{section}}

\def\appendix{\setcounter{section}{0}
        \def\thesection{Appendix.}
        \def\theequation{\Alph{section}.\arabic{equation}}}
\long\def\@makefntext#1{\parindent 0cm\noindent
\hbox to 1em{\hss$^{\@thefnmark}$}#1}
\def\IR{{\hbox{{\rm I}\kern-.2em\hbox{\rm R}}}}
\def\IH{{\hbox{{\rm I}\kern-.2em\hbox{\rm H}}}}
\def\IC{{\ \hbox{{\rm I}\kern-.6em\hbox{\bf C}}}}
\def\IZ{{\hbox{{\rm Z}\kern-.4em\hbox{\rm Z}}}}

\newcommand{\beq}{\begin{equation}}
\newcommand{\eeq}{\end{equation}}
\begin{document}
%
%
%
%
\def\citen#1{%
\edef\@tempa{\@ignspaftercomma,#1, \@end, }
\edef\@tempa{\expandafter\@ignendcommas\@tempa\@end}%
\if@filesw \immediate \write \@auxout {\string \citation {\@tempa}}\fi
\@tempcntb\m@ne \let\@h@ld\relax \let\@citea\@empty
\@for \@citeb:=\@tempa\do {\@cmpresscites}%
\@h@ld}
%
\def\@ignspaftercomma#1, {\ifx\@end#1\@empty\else
   #1,\expandafter\@ignspaftercomma\fi}
\def\@ignendcommas,#1,\@end{#1}
%
%
\def\@cmpresscites{%
 \expandafter\let \expandafter\@B@citeB \csname b@\@citeb \endcsname
 \ifx\@B@citeB\relax 
    \@h@ld\@citea\@tempcntb\m@ne{\bf ?}%
    \@warning {Citation `\@citeb ' on page \thepage \space undefined}%
 \else
    \@tempcnta\@tempcntb \advance\@tempcnta\@ne
    \setbox\z@\hbox\bgroup 
    \ifnum\z@<0\@B@citeB \relax
       \egroup \@tempcntb\@B@citeB \relax
       \else \egroup \@tempcntb\m@ne \fi
    \ifnum\@tempcnta=\@tempcntb 
       \ifx\@h@ld\relax 
          \edef \@h@ld{\@citea\@B@citeB}%
       \else 
          \edef\@h@ld{\hbox{--}\penalty\@highpenalty \@B@citeB}%
       \fi
    \else   
       \@h@ld \@citea \@B@citeB \let\@h@ld\relax
 \fi\fi%
 \let\@citea\@citepunct
}
%
\def\@citepunct{,\penalty\@highpenalty\hskip.13em plus.1em minus.1em}%
%
%
\def\@citex[#1]#2{\@cite{\citen{#2}}{#1}}%
%
%
\def\@cite#1#2{\leavevmode\unskip
  \ifnum\lastpenalty=\z@ \penalty\@highpenalty \fi 
  \ [{\multiply\@highpenalty 3 #1
      \if@tempswa,\penalty\@highpenalty\ #2\fi 
    }]\spacefactor\@m}
\let\nocitecount\relax  
%

\begin{titlepage}
\vspace{.5in}
\begin{flushright}
WATPHYS TH-97/18\\
gr-qc/9808083\\
August 1998\\
\end{flushright}
\vspace{1in}
\begin{center}
{\Large\bf
A Statistical Mechanical Interpretation\\[1ex]
of Black Hole Entropy\\[1ex]\vspace{.08in}
Based on an Orthonormal Frame Action}\\
\vspace{.4in}
R{\sc ichard}~J.~E{\sc pp}\footnote{\it email: epp@avatar.uwaterloo.ca}\\
       {\small\it Department of Physics}\\
       {\small\it University of Waterloo}\\
       {\small\it Waterloo, Ontario N2L 3G1}\\{\small\it Canada}\\
\end{center}

\vspace{.5in}
\begin{center}
\begin{minipage}{5in}
\begin{center}
{\large\bf Abstract}
\end{center}
{\small
Carlip has shown that the entropy of the three-dimensional black hole
has its origin in the statistical mechanics of microscopic states living
at the horizon.  Beginning with a certain orthonormal frame action, and
applying similar methods, I show that an analogous result extends to the
(Euclidean) black hole in any spacetime dimension.  However, this approach
still faces many interesting challenges, both technical and conceptual.
}
\end{minipage}
\end{center}
\end{titlepage}
\addtocounter{footnote}{-1}

\begin{flushleft}
\large\bf Introduction
\end{flushleft}
\noindent
General relativity predicts black holes, which are completely characterized
by just two parameters: their mass and angular momentum.  General relativity
also provides laws of black hole mechanics involving these parameters,
which are strikingly similar to the laws of thermodynamics provided one
makes certain identifications, such as the black hole mass with its
energy, one-quarter the horizon area with its entropy, and so on
\cite{Bard,Beke}.  Hawking showed that this similarity is not accidental:
the laws of black hole mechanics {\em are} the laws of thermodynamics
as applied to a black hole~\cite{Hawk4}.  But the thermodynamics of any
ordinary physical system is only an approximation based on a more fundamental
statistical mechanical description of its microscopic degrees of freedom.
On this last point general relativity seems to be peculiarly silent.
So perhaps general relativity is not a fundamental description of 
gravity and we must search for a deeper theory.  Or---the point of view
taken here---it contains a 
subtlety that when treated carefully will yield a classical
description of certain microscopic degrees of freedom, which when quantized
will yield a highly degenerate set of microstates, and ultimately a
statistical mechanical explanation for black hole entropy.
It is such a subtlety I will discuss in this paper.

The ideas here are inspired by two main themes which stand out in the
literature.  The first is that black hole entropy is intimately connected
with topological considerations, as first emphasized by Gibbons and
Hawking in 1979~\cite{Gibb2}.  Perhaps the strongest statement of this
idea are recent arguments which suggest that the Bekenstein-Hawking
entropy formula ($S=A/4$), generalized to encompass arbitrary topology,
should read~\cite{Libe}
\begin{equation}
\label{eqn:generalized_entropy}
S={\chi\over 8}A\, ,
\end{equation}
where $\chi$ is the Euler number of the black hole, and $A$ is the horizon
area in Planck units.
Topological considerations
will enter the analysis here in at least two places.  Firstly, the analysis
will be based on an orthonormal frame formulation of general relativity, and
a nontrivial spacetime topology is closely linked to the necessary existence
of some set of singular points at which the frame is multivalued.
The action I will be using forces us to excise any such singular
points, introducing boundaries on which, it turns out, the physically relevant
microstates live.  (The general idea of microstates living on boundaries
is introduced more fully beginning in the next paragraph.)
Secondly, some preliminary results I will describe indicate that
a semiclassical counting of the degeneracy of the microstates 
should involve precisely the two parameters $A$ and $\chi$, in a way suggestive
of the general-topology entropy formula (\ref{eqn:generalized_entropy}).

The second inspirational theme is the physically plausible idea that the 
microscopic degrees of freedom, if they exist at all within the context of 
general relativity, will likely be associated in some way with the event 
horizon (thought of as a `boundary' of sorts); after all, the entropy is 
proportional to the area of the event horizon.  
The first concrete realization of this idea occured in the context of the
(2+1)-dimensional black hole discovered by Ba\~{n}ados, Teitelboim, and
Zanelli (BTZ)~\cite{BTZ}, when Carlip showed that its entropy can be understood
in terms of the statistical mechanics of certain microstates living at
a boundary defined by apparent horizon boundary conditions~\cite{Carl}.
Closely related is the independent work by Balachandran, Chandar, and
Momen~\cite{Bala}.
The basic idea in Carlip's approach is to use the fact that in 
three dimensions general relativity can be formulated as a (Chern-Simons)
gauge theory~\cite{Achu,Witt2}, and the presence of a boundary breaks the
gauge symmetry leading to a Wess-Zumino-Witten (WZW) boundary action 
\cite{Witt1,Elit} for ``would-be gauge'' degrees of freedom now promoted
to physical gravitational boundary degrees of freedom.  Enforcing
apparent horizon boundary conditions and quantizing the resulting boundary
theory leads to a set of microstates whose degeneracy correctly accounts
for the BTZ black hole entropy~\cite{Carl,Carl4,Carl3}.\footnote{For a recent
review, and discussion of some unresolved questions in this approach,
see Ref.~\cite{Carl5}.} 
But this approach seems to
rely on a gauge-theoretic formulation of general relativity, 
making it specific to gravity in three dimensions.  
I will show that this
might unnecessarily be asking too much: all we require is a formulation
of general relativity which has a gauge symmetry that is broken when a
boundary is present.  
(I.e., we do not require such gauge transformations to be equivalent on-shell
to diffeomorphisms, as is the case in three-dimensional gravity~\cite{Witt2}.)
I will introduce such a formulation and show how
it opens the door to extending the main idea of Carlip's approach to
black holes in any spacetime dimension.

Baez {\it et al}~\cite{Baez} have also found evidence for boundary degrees
of freedom, this work in the context of the loop variables 
approach to quantum gravity.  More recently, Ashtekar {\it et al}~\cite{Asht1}
have introduced a `black hole sector' of non-perturbative canonical quantum
gravity in which the quantum black hole degrees of freedom (i.e., the
microstates) are described by a Chern-Simons field theory on the horizon
(see also Refs.~\cite{Asht2,Husa}).  They show that, for the case of
a large non-rotating black hole, counting the degeneracy of these microstates
leads to an entropy proportional to the area $A$ divided by the so-called
Immirzi parameter.  An appropriate choice of this parameter then yields
the Bekenstein-Hawking entropy.  (A somewhat analogous undetermined parameter
appears in my analysis---I will comment on this in Section~6.)

Progress towards understanding black hole microstates has also been made
on the string theory front.  (A review of some of the original ideas
can be found in Ref.~\cite{Horo}.)  For an example immediately
relevant here, Sfetsos and Skenderis~\cite{Sfet} have shown that the
ordinary (four-dimensional) Schwarzschild black hole is U-dual to the
(three-dimensional) BTZ black hole so one can count its microstates
by counting the BTZ black hole microstates following Carlip's approach,
and they find the correct Bekenstein-Hawking entropy.
We will encounter numerous other hints that the approach described
here may have hidden connections with the string theory
approach to black hole microstates.

The main point of the previous three paragraphs is to emphasize that the
idea that microstates can be somehow associated with boundaries, in
particular the event horizon, is a viable one.  My point of view then
is to simply do general relativity on a manifold with boundary and see
what happens.
Mathematically this certainly opens up a wealth of possibilities, and
whether or not the results are physically meaningful---in particular,
{\em can} we think of the event horizon as a boundary, and if so, 
precisely {\em how}---can be decided after the analysis.
Nevertheless, at this point I can raise some anticipated objections.
For instance, one might be uncomfortable thinking of the horizon as
a `tangible' boundary, or might wonder, ``What about the microstates
which would presumably reside also at the boundary at infinity?''
Such concerns will be neatly dealt with in the approach I will now
outline.

In Section~1 we will begin with a $d$-dimensional `bare' manifold, $M$, 
with boundary $\partial M$,
not being at all specific about the physical nature of this boundary.
We will then endow this manifold with an orthonormal frame and 
suppose that, fundamentally, this frame (rather than the metric) encodes
the gravitational degrees of freedom.  This introduces additional degrees
of freedom, namely local frame rotations, which we will want to be pure
gauge at least in the interior of $M$, but not necessarily on its boundary.
We will then introduce
a very natural action, first order in derivatives of the frame field, which
is gauge invariant except for a boundary term that breaks
the gauge symmetry on $\partial M$.  Applying the main idea of Carlip's 
program then leads to a certain boundary action for
``would-be gauge'' frame rotation degrees of freedom now promoted to physical
gravitational boundary degrees of freedom.  
In the remainder of the paper I will discuss the nature and interpretation
of this boundary theory, restricting attention mainly to the case of Euclidean
black holes.

To begin with, in Section~2 I will describe where the microstates live,
and how this question is related to singularities (i.e., multivaluedness)
of the frame.  Section~3 is devoted to the action principle, boundary
conditions, and Euler-Lagrange equations.  Sections~4 and 5 contain a
description of the boundary theory phase space and its relation to
a higher dimensional generalization of Ka\v{c}-Moody and Virasoro algebras.
Finally, in Section~6 I will address the issues of quantization and the
counting of the degeneracy of the microstates.  Unfortunately, at the 
technical level this is a rather formidable task, and some detailed results
are worked out in the three spacetime dimensions case only.
For the higher dimensional cases it is perhaps worth noting at this point that 
the mathematics involved is reminiscent of that used in the study of $p$-branes
and non-perturbative effects in string theory.

\section{The Proposed Action}
\label{sec:2}
\noindent
In a metric formulation of gravity the gravitational degrees of freedom
are encoded in the metric, $g$, and the standard action, second order in
derivatives of the metric, is
\begin{equation}
\label{eqn:standard_action}
I^{(2)}[g]={1\over{2\kappa}}\int_{M}\epsilon\, R\, .
\end{equation}
Here $M$ is the 
spacetime manifold, on which $g$ induces a volume form, $\epsilon$, and
scalar curvature $R$.  In $d$ spacetime dimensions $\kappa$ is a numerical
factor times $(l_P)^{d-2}$, where $l_P$ is the Planck length.  It sets the
scale of the action, 
and enters any quantum statement following from this action.
If the manifold has a boundary, $\partial M$, 
we must consider what boundary term (if any) 
to add to this action, 
knowing that the choice we make is important to
the physics we are trying to describe.
For instance, one might add a boundary term to accommodate certain prescribed
boundary conditions, such as microcanonical boundary conditions 
\cite{Brow2}, but in any case I wish 
to emphasize that this boundary term is usually put in by hand.

Now, for reasons which will soon become apparent, let us switch to a 
formulation of gravity in which the gravitational degrees of freedom are
encoded, not in the metric, but rather its `square root': the one-form
fields $e^a$, $a=0,1,\ldots ,d-1$, which at any given spacetime point
represent, physically, the orthonormal reference frame associated with
an observer at that point.  It is a simple exercise to show that
\begin{equation}
\label{eqn:epsilon_R}
\epsilon\, R=
\omega^{a}_{\;\;b}\wedge\omega^{b}_{\;\;c}\wedge\epsilon^{c}_{\;\;a}
-d\, (\omega^{a}_{\;\;b}\wedge\epsilon^{b}_{\;\;a})\, ,
\end{equation}
where (and this is important) the connection one-form, $\omega^{a}_{\;\;b}$, is
not an independent field, but rather it is determined uniquely
from a given frame field 
through the usual metricity ($\omega^{ab}=-\omega^{ba}$) and no-torsion
($de^{a}+\omega^{a}_{\;\;b}\wedge e^{b}=0$) conditions.  We have defined
the $(d-2)$-form $\epsilon_{ab}=i_{e_b}i_{e_a}\epsilon$, where the 
vector fields $e_a$, dual to the one-form fields
$e^a$, are defined by the relations
$i_{e_a}e^{b}=\delta^{b}_{a}$, and frame indices 
are raised and lowered with the
matrix $\eta^{ab}=\eta_{ab}$, diagonal with $s$ $-1$'s and $(d-s)$ $+1$'s,
reflecting the signature of the spacetime metric.  The frame rotation group
is $G=SO(d-s,s)$, with Lie algebra $g=so(d-s,s)$.\footnote{Taken in context,
there should be no confusion between this $g$ and the spacetime metric, $g$.}

It is useful to introduce the notation
\begin{equation}
\label{eqn:EW}
E={\textstyle\rm matrix}\, (-\epsilon^{a}_{\;\;b})\, , \;\;\;
W={\textstyle\rm matrix}\, (\omega^{a}_{\;\;b})\, ,
\end{equation}
with $a$ ($b$) a row (column) index, and to think of $E$ as a $g$-valued
$(d-2)$-form, and $W$ as a $g$-valued one-form.  The gravitational action
I am advocating is first order in derivatives of the frame field
and comes from simply dropping the total divergence on the right hand side
of (\ref{eqn:epsilon_R}):
\begin{equation}
\label{eqn:proposed_action}
I^{(1)}[e]=-{1\over{2\kappa}}\int_{M}{\textstyle\rm Tr}\,
(W\wedge W\wedge E)=I^{(2)}[g(e)]+I_{\partial M}[e]\, ,\;\;\;
I_{\partial M}[e]=-{1\over{2\kappa}}\int_{\partial M}
{\textstyle\rm Tr}\, (W\wedge E)\, ,
\end{equation}
where Tr denotes matrix trace.  Notice that $I^{(1)}[e]$ suggests a boundary
term, namely $I_{\partial M}[e]$, that is to be added to the bulk action
in (\ref{eqn:standard_action}).  My point of view is to suppose that
this action is the correct action to describe gravity in a spacetime of 
any dimension, signature, and topology, without the need to augment it with
any additional boundary terms put in by hand.  It is our job now to simply
analyze it and extract the physics it contains.

This action is not really new.
In the $d=4$ Lorentzian case a bit of index
manipulation shows that $I^{(1)}[e]$ is the same action proposed by
Goldberg~\cite{Gold}, which he uses to derive a formalism equivalent to
the Ashtekar-variables approach~\cite{Asht}. However, Goldberg's
derivation of $I^{(1)}[e]$ is different, and is based on the so-called
Sparling-Thirring forms.  These forms are interesting 
in themselves because
they satisfy a relationship that expresses the Einstein tensor in terms of
an energy-momentum pseudotensor and a superpotential~\cite{Gold}.  
Related to this
fact is work by Lau~\cite{Lau1,Lau2} who
uses this Goldberg action to derive an Ashtekar-variables reformulation of the
metric theory of quasilocal stress-energy-momentum originally due to 
Brown and York~\cite{Brow1}.  So it would seem that $I^{(1)}[e]$ is
particularly Ashtekar-variables-friendly, but 
I will not explore this aspect of the formalism here.
Finally, $I^{(1)}[e]$ is similar to the first order
action originally proposed by Einstein~\cite{York}, but differs from 
it by a tetrad expression that is not expressible in terms of the metric
itself.  The main contribution of this paper is the recognition that 
$I_{\partial M}[e]$ contains a boundary action describing new physical
gravitational degrees of freedom living on $\partial M$, and the derivation
of what I think are interesting and significant consequences of this fact.

Regarding symmetries, we first observe that the proposed action is obviously
invariant under diffeomorphisms which preserve the boundary, and no motivation
will present itself to contemplate diffeomorphisms outside of this restriction.
Secondly, the proposed action is invariant under local frame rotations except
at the boundary, where the undifferentiated $W$ in $I_{\partial M}[e]$ breaks
this symmetry.  A broken symmetry such as this might at first be thought of as
a defect of the action, but it is precisely this
feature which allows us to apply Carlip's program~\cite{Carl,Carl4,Carl3} 
to arrive
at gravitational boundary degrees of freedom, which will eventually lead to
the microstates I am suggesting might be responsible for black hole entropy.
And this is the principal reason for going to an orthonormal frame (rather
than metric) formulation of gravity.

So, following in the spirit of Carlip's work, let us parametrize the frame
as $e^{a}=U^{a}_{\;\;b}\hat{e}^{b}$, where $\hat{e}^{b}$ is a gauge-fixed
frame (with corresponding gauge-fixed connection, 
$\hat{\omega}^{a}_{\;\;b}$), and all frame rotation degrees of freedom are
included in $U={\textstyle\rm matrix}\, (U^{a}_{\;\;b})\in G$.  In terms of
this matrix notation the parametrization reads
\begin{equation}
\label{eqn:parametrized_EW}
E=U\!\hat{E}U^{-1}\, ,\;\;\;
W=U\hat{W}U^{-1}+UdU^{-1}\, ,
\end{equation}
and the action accordingly splits:
\begin{equation}
\label{eqn:split_action}
I^{(1)}[e]=I^{(1)}[\hat{e}]+I_{B}[U;\hat{e}]\, ,
\end{equation}
into a gauge-fixed action, $I^{(1)}[\hat{e}]$, 
plus (what I shall interpret as) a boundary
action:
\begin{equation}
\label{eqn:boundary_action}
I_{B}[U;\hat{e}]={1\over{2\kappa}}\int_{\partial M}
{\textstyle\rm Tr}\, (U^{-1}dU\wedge \hat{E})\, .
\end{equation}
I will analyze these two objects in turn.

\section{The Gauge-Fixed Action, $I^{(1)}[\hat{e}]$}
\label{sec:3}
\noindent
The gauge-fixed action depends on the choice we make for the gauge-fixed
frame on $\partial M$.  But of course all choices are equivalent, at least
at the classical level, since a different choice can be absorbed into a change
of variables of the $U$ degrees of freedom.  So the choice is merely a matter
of convenience, rather than `putting in physics by hand.'  It turns out
that a convenient choice is to gauge-fix one leg of the vector frame, denoted
as $\hat{e}_{\perp}$, to equal the unit normal, $n$, everywhere on 
$\partial M$.\footnote{$I^{(1)}[\hat{e}]$ contains some interesting physics 
when the boundary
is null, but since I have not yet analyzed the boundary action for this case,
which in my principal focus here, we will restrict ourselves to the case where
$\partial M$ is nowhere null.}  A short calculation reveals that with this
choice the gauge-fixed action is
\begin{equation}
\label{eqn:gauge-fixed_action}
I^{(1)}[\hat{e}]={1\over{2\kappa}}\int_{M}\epsilon\, R+
\pi_{n}{1\over\kappa}\int_{\partial M}\epsilon_{\partial M}\, K\, ,
\end{equation}
where $K$ is the trace of the extrinsic curvature of the boundary, 
$\epsilon_{\partial M}=i_{n}\epsilon$ is the boundary volume form, and
$\pi_{n}=n\cdot n=\pm 1$, the sign depending on the relative signatures of
the metrics on $M$ and $\partial M$.  This result is independent of how
we rotate the remaining legs of the gauge-fixed vector frame 
($\hat{e}_{a}$, $a\not =\perp$) in the tangent space of $\partial M$.  We
see that $I^{(1)}[\hat{e}]$ is just the standard Einstein-Hilbert action
appropriate for holding fixed, in the action principle, the boundary
$(d-1)$-geometry~\cite{York}.  A similar result for the Goldberg action
was first shown by Lau~\cite{Lau2}.

But this is not the end of the story.  For example, suppose we consider a
Lorentzian spacetime in which a spacelike section of $\partial M$
(with unit normal $u$) joins a timelike section (with unit normal $n$) in
a $(d-2)$-dimensional `corner,' $\cal C$.  If we gauge-fix an appropriate
leg of the vector frame on each section as described in the previous paragraph
(say $\hat{e}_{0}=u$ and $\hat{e}_{1}=n$ on the respective sections) we pick
up a trace $K$ term from each, as in (\ref{eqn:gauge-fixed_action}), but
since on $\cal C$ $n\cdot u\not =0$, in general, the frame is double-valued
on $\cal C$.  It is easy to render it single-valued simply by introducing
a finite frame rotation (in this case a boost) in an infinitesimal neighborhood
of $\cal C$, which has the effect of augmenting the right hand
side of (\ref{eqn:gauge-fixed_action}) with an additional `corner term' of
the form
\begin{equation}
\label{eqn:corner_term}
{1\over\kappa}\int_{\cal C}\epsilon_{\cal C}\,
{\textstyle\rm arcsinh}\, (n\cdot u)\, ,
\end{equation}
where $\epsilon_{\cal C}$ is the corner volume form.  
Again, in the context of the
Goldberg action, this sort of calculation was first done by Lau~\cite{Lau2}
so I will not reproduce a similar calculation here.
It is easy to demonstrate that, when $\partial M$ has corners, precisely such
corner terms are required to make the Einstein-Hilbert action well defined.
This has been known for some time~\cite{Hart,Hayw} (see also Ref.~\cite{Bril}),
and it is satisfying that these corner terms arise {\it naturally} in 
$I^{(1)}[\hat{e}]$ simply by demanding single-valuedness of the frame.
Indeed, this is one reason I am suggesting that $I^{(1)}[e]$ might be 
`universal'---recall my comment immediately following 
(\ref{eqn:proposed_action}).  Furthermore, this nice property of
$I^{(1)}[\hat{e}]$ provides motivation to take seriously the other piece
of $I^{(1)}[e]$, namely $I_{B}[U;\hat{e}]$ in (\ref{eqn:split_action}),
which will be the principal focus of this paper.

Now let us turn our attention to Euclidean black hole spacetimes, assuming a
topology $M=R^{2}\times S^{d-2}$ with boundary 
$\partial M=S^{1}\times S^{d-2}$ at infinity.  In this case $\partial M$
has no corners and there appears to be no problem in gauge-fixing, say,
$\hat{e}_{0}=n$ (and $\hat{e}_{1}=t$, a Euclidean time flow unit vector
tangent to $\partial M$), to obtain (\ref{eqn:gauge-fixed_action}).  However,
now a subtlety of a different sort arises: This choice of gauge is analogous
to using the frame $\hat{e}_{0}=\partial/\partial r$, 
$\hat{e}_{1}=(1/r)\partial/\partial\phi$ on a flat disk with polar coordinates
($r,\phi$), the disk being analogous to the $R^{2}$ sector of $M$.  On the
boundary ($r=1$) this is all right, but the frame is multivalued at the
origin, and no other choice of frame in the interior (smoothly connected to
our choice on the boundary) can avoid such a point of multivaluedness.
Now,  
although this disk example falls short of a proof, it seems highly plausible
that a similar problem obtains in the black hole case, where this type
of multivaluedness must exist on some set of points in the interior of $M$, say 
the bifurcation sphere $\{0\}\times S^{d-2}$.  
(And even if one could prove that this is not {\it necessarily} the case,
such a choice of frame is certainly possible, and arguably natural.)
One might wonder, ``So what?
Such frames are used all the time, and their multivaluedness is no more
problematic than a mere coordinate singularity.''  I argue that this is
not so:  First, we learned in the Lorentzian case
that demanding single-valuedness of the frame is necessary to obtain the
correct corner terms.  And secondly, if the proposed action is to be
taken seriously as a description of gravity based not on the metric, but
on observers' frames, it seems physically unreasonable to allow more than one
frame attached to a given spacetime point.

I think these arguments provide sound motivation for excising the bifurcation
sphere from the spacetime, which obviously removes the 
multivaluedness problem,
and is the simplest (but not the only)\footnote{Returning to the disk
example, it is also possible to begin
with the choice $\hat{e}_{0}=\partial/\partial x$, 
$\hat{e}_{1}=\partial/\partial y$ in Cartesian ($x$,$y$) coordinates, so that
at the point (1,0) on the boundary $\hat{e}_{0}=n$.  Then as one circles
counterclockwise around the boundary the frame is rotated 
such that $\hat{e}_{0}$
remains in the normal direction.  Through the last infinitesimal segment 
to complete the circuit a clockwise $2\pi$ rotation is executed to return the
frame to its original orientation.  It is clear that this operation on the 
boundary can be continued smoothly into the interior, and the resulting frame
is single-valued everywhere.  The $2\pi$ rotation in 
an infinitesimal neighborhood of
the boundary point (1,0) produces a `topological' (as opposed to corner)
term, which in the black hole case 
turns out to equal $-2\pi/\kappa$ times the volume of
the $(d-2)$-sphere fiber at the point (1,0).  This seems unsatisfactory
in that our starting point, (1,0), becomes a preferred point (unless
the spacetime admits a Euclidean time Killing vector field on $\partial M$).
For this, and other reasons, we will adhere to the solution of the
multivaluedness problem presented in the main text 
(exception: see footnote~5).} 
solution.  This subtlety is important because
now the spacetime becomes an annulus cross $S^{d-2}$, with both an outer
and an inner $S^{1}\times S^{d-2}$ boundary, the latter bounding a `thickened
bifurcation sphere.' This means there will be boundary degrees of 
freedom (discussed in the next section) 
on both outer and inner boundaries, which on quantization will give
rise to two independent sets of microstates.  If $\Gamma_o$ and $\Gamma_i$ 
denote respectively the degeneracies of these microstates, then the total
degeneracy of microstates will be 
the product, $\Gamma_o \Gamma_i$.  And according
to statistical mechanics, the total entropy will be $\ln\Gamma_o +\ln\Gamma_i$
(with Boltzmann's constant set to one).  
If the boundary theory I am suggesting here works at all we would expect
$\ln\Gamma$ to be proportional to the volume of a $(d-2)$-sphere
cross section of the boundary, which means that
$\ln\Gamma_o$ will be divergent.  However, it is well known that the action in 
(\ref{eqn:gauge-fixed_action}) is also divergent when evaluated on
a Euclidean black hole spacetime (due to the trace $K$ term integrated over
the outer boundary at infinity), 
and that the {\em physical} action is obtained by
subtracting the same action evaluated on a suitable vacuum spacetime, yielding
a finite result~\cite{Brow1,Brow2,Brow3,Hawk,Gibb}.
The analogous regularization procedure here
involves subtracting the entropy of the vacuum spacetime.  This will consist
of the {\em same} $\ln\Gamma_o$ term (since, by definition, a ``suitable''
vacuum spacetime has the same intrinsic outer boundary geometry as the black
hole), but no $\ln\Gamma_i$ term.  (We may still need to excise some set of
points from the interior of the vacuum spacetime
to obtain a single-valued frame, but a
little thought shows that the boundary theory 
resulting from this excision is trivial---I will elaborate on this in 
Section~6.)
So the physical entropy is just $\ln\Gamma_i$, precisely what we want, or 
might expect.

This result may be viewed as a direct response to a speculation by Carlip
and Teitelboim (the last sentence in Ref.~\cite{Carl2}) 
in which they suggest that the numerical
factor relating the entropy of a Euclidean black hole to the area of its
bifurcation sphere might have its origin in microstates living on 
the boundary of a 
thickened bifurcation sphere.
We will see later (equation (\ref{eqn:main_result_entropy})) 
that this numerical
factor is deeply connected with the nature of the quantized boundary theory.

Furthermore, consider the following.  It is
well known that the topology of a manifold is intimately connected with
the index of certain vector (or one-form) fields with isolated singularities
in the manifold and on its boundary~\cite{Will,Arno}.  Insofar as the
vector (or one-form) frame field 
(rather than the metric) is the fundamental object out of which
$I^{(1)}[e]$ is constructed, and it is precisely a certain type of
isolated singularity (the multivaluedness) of this frame field which 
motivated excising the bifurcation sphere, leading to the above $\ln\Gamma_i$
contribution, it is tempting to speculate that this contribution is
really of topological origin.  What I mean is that this
appears to be an example of a general mechanism whereby
$I^{(1)}[e]$ encodes physical consequences of the topology of the 
spacetime.
Stated even more strongly, the entropy of a black hole is a topological
phenomenon.  This point of view is advocated by Gibbons and 
Hawking~\cite{Gibb2} 
(see also Refs.~\cite{Libe,Hawk,Hawk2,Bana}),
but notice that what is suggested here is something more: it is not
merely the Bekenstein-Hawking  entropy (one-quarter the horizon area) we are
discussing, but rather the quantum microstates responsible for that entropy.

The `excision principle' introduced above is crucial to the physics we
are considering.  I have given a physical justification for it, now let
me provide a more mathematical one.  To be concrete let us consider the
example of the Euclidean-Schwarzschild spacetime (for simplicity
without a conical singularity), which is topologically $R^{2}\times S^{2}$.
It is easy to verify 
for this metric that equation (\ref{eqn:epsilon_R}) is satisfied at every point
in $M$.  Now when we integrate this equation over $M$ the left hand side
is of course zero, and the right hand side gives a volume term (call it $V$)
plus a surface term (call it $S$), the latter an integration over $\partial M$
(proportional to $I_{\partial M}[e]$).  We require $V+S=0$, but the question
arises, ``What is $\partial M$?''  If we assume that the 
Euclidean-Schwarzschild
spacetime has a boundary only at infinity we get a wrong result: $V+S\not=0$.
Perhaps somewhat surprisingly, it turns out that to get the correct result
we must assume that $\partial M$ consists of the usual boundary at infinity
{\em plus} the boundary (call it $\partial M_{*}$) of the following
suitably `thickened'  set of
points: all points comprising the bifurcation sphere,
and the north and south poles of the two-sphere at each point of $R^2$.
What is happening here?  Having chosen one leg of the tetrad in the radial
direction and another in the Euclidean time direction, this pair of legs is
multivalued on the bifurcation sphere, as discussed previously.  Furthermore,
the two-sphere is not parallelizable: if the remaining two legs of the tetrad
are chosen in the $\partial/\partial\theta$ and $\partial/\partial\phi$
directions, respectively (where $\theta$ and $\phi$ denote standard spherical
coordinates), then this pair of legs is multivalued at the two poles of
the sphere.  The lesson from this example is that the action $I^{(1)}[e]$ 
{\em tells us} what $\partial M$ must be: in order for the
${\rm Tr}\,(W\wedge W\wedge E)$ volume integral in (\ref{eqn:proposed_action})
to equal $I^{(2)}[g(e)]+I_{\partial M}[e]$ we are {\em forced} to excise
from $M$ all points at which the frame is multivalued.  This set of points,
being of measure zero, does not affect the volume integral, but does
affect the boundary term $I_{\partial M}[e]$.  As a consequence, microstates 
live on all points of $\partial M_{*}$ (as well as on the boundary at infinity,
but as argued above, regularization renders these non-physical).  
In the Euclidean-Schwarzschild
example it is easy to show that the boundary theory on the portion of 
$\partial M_{*}$ corresponding to excision of the poles of $S^2$ is trivial,
but that on the portion of $\partial M_{*}$ which is the boundary of the 
thickened bifurcation sphere is not, and it is here that the physically
relevant microstates in this example live.  A similar phenomenon occurs in
each of several examples I have studied, and thus it appears that it may
be generic.

Furthermore, the usual value of the regularized on-shell Euclidean
action (taking into account only the boundary at infinity) is changed 
when we take into account $\partial M_{*}$, and consequently so is the 
thermodynamical entropy calculated from the zero-order partition function.
(Generically this is so, but not, it turns out, in the 
Euclidean-Schwarzschild case, due
to some `miraculous' cancellations.)  It has been emphasized by Brill
and Hayward that the action for a spacetime is not necessarily invariant
under topological identifications of isometric surfaces on its 
boundary; a finite action is associated with certain identification 
surfaces~\cite{Bril}.  Their analysis involves precisely the type of
corner terms discussed earlier, and it seems likely that their findings
are closely connected to what is happening here with $I^{(1)}[e]$.
  
In summary, let me reemphasize that necessary singularities
(multivaluedness) of the vector or one-form frame are associated with a
nontrivial topology of $M$, 
and $I^{(1)}[e]$ has a remarkable built-in sensitivity to this measure of
topology, which manifests itself in an excision principle, and which in turn
dictates the location where the physically relevant microstates 
live.\footnote{Recently Strominger~\cite{Stro} has given a new microscopic
derivation of the BTZ black hole entropy based on the observation
\cite{Brow4} that the asymptotic symmetry group of anti-de Sitter space
is infinite dimensional, generated by (two copies of) the Virasoro algebra.
His derivation suggests that the microscopic degrees of freedom might reside
at infinity, rather than at the horizon.  (Or perhaps in this special case
both possibilities can be shown to be equivalent.)  
This may be a good example in which to test the alternative to the excision
principle mentioned in footnote~4, but I will not do so here.}
Thus, one might say that entropy is a topological phenomenon not only at the
level of thermodynamics, but also at the level of statistical mechanics.

\section{The Action Principle}
\label{sec:4}
\noindent
For clarity of exposition in this section we shall restrict ourselves to
the `regular' sections of $\partial M$, ignoring corners if any, and,
where necessary, we assume that we have excised a suitable set of points
from $M$ such that, as discussed in the previous section, we can gauge-fix
$\hat{e}_{\perp}=n$ everywhere on $\partial M$.  The $(d-1)$ vectors
$\hat{e}_{i}$ ($i$ running over the values of $a$, $a\not=\perp$) are
everywhere tangent to $\partial M$.  That the vector frame $\hat{e}_{a}$ is
gauge-fixed means that, when computing variations, any occurence of the 
combination
$\hat{e}_{[a}\cdot\delta\hat{e}_{b]}$ is set to zero, as it represents a
rotation degree of freedom already accounted for in $U^{a}_{\;\;b}$.  
Finally, let $\mu,\nu,\ldots$ denote 
spacetime tensor indices referred to a set of local
coordinates in $M$ (or their restriction to $\partial M$), which are raised
and lowered with the spacetime metric $g_{\mu\nu}$.

Varying the action $I^{(1)}[e]$ in (\ref{eqn:split_action}) with respect to
$\hat{e}$ and $U$ we obtain
\begin{eqnarray}
\delta I^{(1)}[e]&=&
{1\over 2\kappa}\int_{M}\epsilon\, 
(R_{\mu\nu}-{1\over 2}R\, g_{\mu\nu})\, \delta g^{\mu\nu}(\hat{e})-
{1\over 2\kappa}\int_{\partial M}{\textstyle\rm Tr}\,(dE\,\delta U U^{-1})
\nonumber \\
& & \mbox{} +
\pi_{n}{1\over 2\kappa}\int_{\partial M}\epsilon_{\partial M}\,
(\Pi_{\mu\nu}+\kappa T_{\mu\nu})\, \delta h^{\mu\nu}(\hat{e})\, .
\label{eqn:variation_of_action}
\end{eqnarray}
So for the action to be extremized we require, first, that the metric
$g_{\mu\nu}(\hat{e})$, constructed out of $\hat{e}^{a}$, satisfy the vacuum
Einstein equations at every point in $M$.\footnote{Notice that excising a set
of points from $M$ naturally
allows for the possibility of introducing conical
singularities at these points without affecting the bulk term in
(\ref{eqn:variation_of_action}), but I will not discuss this generalization
here.}  
The second integral comes from
varying the boundary action, $I_{B}[U;\hat{e}]$, with respect to $U$,
and it will vanish if and only if the boundary degrees of freedom, $U$, satisfy
the Euler-Lagrange equations
\begin{equation}
\label{eqn:Euler-Lagrange_equations}
0=dE\downarrow_{\partial M}=d\, (U\!\hat{E}U^{-1})\downarrow_{\partial M}\, ,
\end{equation}
where the symbol $\downarrow_{\partial M}$ denotes pullback 
of forms to $\partial M$.
Recall that $\hat{E}$ is a $g$-valued $(d-2)$-form (constructed out of the
gauge-fixed frame); $E=U\!\hat{E}U^{-1}$ is its orbit under the action of
$U\in G$.  These boundary equations for $U$ are deceptively 
simple-looking---in fact they are highly nontrivial.  I will not present any
results of the analysis of them in this paper, except for some brief comments 
in Section~\ref{sec:5}. 

In the third integral on the right hand side of 
(\ref{eqn:variation_of_action}), $h_{\mu\nu}(\hat{e})$ is the induced boundary
metric, constructed out of $\hat{e}^{i}$.  
$\Pi_{\mu\nu}=K_{\mu\nu}-Kh_{\mu\nu}$ is the usual gravitational momentum
canonically conjugate to $h^{\mu\nu}$ 
(and $K_{\mu\nu}=h_{\mu}^{\;\;\rho}\nabla_{\rho}n_{\nu}$ is the extrinsic
curvature of the boundary).  It is a fact that the boundary action 
$I_{B}[U;\hat{e}]$ depends on $\hat{e}$ only through $h_{\mu\nu}(\hat{e})$.
So varying $I_{B}[U;\hat{e}]$ with respect to $\hat{e}$ amounts to calculating
the energy-momentum tensor for the $U$ degrees of freedom, the result being:
\begin{equation}
\label{eqn:energy-momentum_tensor}
T^{\mu\nu}={1\over\kappa}[\eta^{ij}\eta^{kl}-\eta^{ik}\eta^{jl}]\,
i_{\hat{e}_{i}}(U^{-1}dU)_{j\perp}\, 
\hat{e}_{k}^{\;\;(\mu}\hat{e}_{l}^{\;\;\nu )}\, ,
\end{equation}
which has the standard `$p\partial q$' form. 
(Since the boundary action is linear in first derivatives of $U$, $U$ is 
already a phase space variable.)
As remarked earlier, the action $I^{(1)}[e]$ is
invariant under diffeomorphisms which preserve $\partial M$.  This means that
on-shell we must have $D_{\nu}(\Pi^{\mu\nu}+\kappa T^{\mu\nu})=0$, where 
$D_{\nu}$ is the covariant derivative operator
induced on $\partial M$.  But we already know that 
$D_{\nu}\Pi^{\mu\nu}=0$ on-shell---this is just the momentum constraint
of general relativity on the $(d-1)$-surface $\partial M$.  So we
must have $D_{\nu}T^{\mu\nu}=0$ on-shell.
Indeed, this can be explicitly shown to follow
from the boundary equations (\ref{eqn:Euler-Lagrange_equations}) and the
above definition of $T^{\mu\nu}$.  So when $h_{\mu\nu}(\hat{e})$ admits
Killing vectors one can construct corresponding 
boundary conserved charges, such as
the total energy, momentum, or angular momentum corresponding to a given
solution $U$.

Now there are two ways to interpret the third integral on the right hand side
of (\ref{eqn:variation_of_action}) with regard to extremizing the action.  The
first is to suppose, as in the standard Einstein-Hilbert action, that one
holds fixed the boundary $(d-1)$-geometry, i.e., $h_{\mu\nu}(\hat{e})$ up to
diffeomorphisms.  
(In this case the `total momentum' $\Pi_{\mu\nu}+\kappa T_{\mu\nu}$ is not
restricted on $\partial M$.)
Following in the spirit of Carlip's 
work~\cite{Carl,Carl4,Carl3}
the following statistical mechanical interpretation of the boundary theory
can then be given.  One solves the vacuum Einstein equations for 
$g_{\mu\nu}(\hat{e})$ in $M$ to obtain a `macrostate,' for instance a
black hole solution.  Information about this macrostate, for example
the black hole mass and angular momentum, enters the boundary action
$I_{B}[U;\hat{e}]$ through $\hat{e}$ (or equivalently, the boundary
metric $h_{\mu\nu}(\hat{e})$). $h_{\mu\nu}(\hat{e})$ is thought of as a
fixed background metric on $\partial M$, whereas the boundary degrees of
freedom, $U$, are dynamical, and, on quantization, lead to the microstates
responsible for the entropy of this macrostate.  As for a physical
interpretation of the boundary degrees of freedom, I think the following one
is both simple and compelling:  If there is {\it any} sense in which a 
boundary in spacetime is physically meaningful\footnote{See section III of
Ref.~\cite{Carl3} for further discussion on this point.} 
(event horizon, apparent horizon, 
thickened bifurcation sphere~\cite{Carl2},
membrane model of horizon~\cite{Magg}, 
black hole complementarity approach~\cite{Suss}, 
't~Hooft's ``brick wall'' approach~\cite{tHoo} 
(and later refinements~\cite{Myer}), 
black hole in a box, e.g.~\cite{Brow2}, 
quasilocal quantities associated with spatially bounded regions,
e.g.~\cite{Brow1}, 
boundary at infinity, and so on), then the orientation
of an observer's frame relative to this boundary (encoded in $U$) is
{\em likewise} physically meaningful.

A second way to extremize the action in (\ref{eqn:variation_of_action}) is
what might be called a `no boundary condition proposal.'  One allows
$h_{\mu\nu}(\hat{e})$ to vary freely on the boundary (hence no boundary
conditions), in which case one is forced to make the identification
\begin{equation}
\label{eqn:T=Pi}
T^{\mu\nu}=-{1\over\kappa}\Pi^{\mu\nu}\, .
\end{equation}
This identification seems natural in light of the work by 
Brown and York on quasilocal quantities associated with
spatially bounded regions in general relativity, wherein they define 
essentially the
right hand side of (\ref{eqn:T=Pi}) as the boundary gravitational 
energy-momentum tensor~\cite{Brow1}.  
So the identification (\ref{eqn:T=Pi}) in some
way reflects a coupling between bulk ($g_{\mu\nu}(\hat{e})$) and 
boundary ($U$) degrees of freedom.  (In a different context, a coupling 
of this sort is discussed in Ref.~\cite{Bala}.)
  
I have already mentioned
that $D_{\nu}T^{\mu\nu}=0$ follows from the boundary equations
(\ref{eqn:Euler-Lagrange_equations}), and so under the identification
(\ref{eqn:T=Pi}), the boundary equations imply the momentum constraints of
general relativity.  (Which remains true if we replace $\partial M$ in our
analysis with any $(d-1)$-dimensional now initial-value surface in
$M$.)  Applied in the simplest case, the 
($d=3$) Euclidean BTZ black hole~\cite{BTZ} (for which, of course, 
we augment $I^{(1)}[e]$ with a cosmological term), 
it turns out that 
the $U$ boundary equations also imply the Hamiltonian constraint of 
general relativity, and there are indications that this may
be true in general.  Even more surprising, these same indications 
hint that the
converse may also be true: under the identification (\ref{eqn:T=Pi}) our
boundary theory for $U$ may be {\em equivalent} to the initial-value 
constraints of general relativity.   
To appreciate the potential significance of this, bear in mind that
essentially the full content of general relativity is encoded in its
initial-value constraints, in that, given a good set of initial data on a 
Cauchy slice, the full spacetime is generated canonically by these same
constraints.  And to the extent that our boundary theory for $U$ can be
quantized, and in principle, at least, solved (see Section~\ref{sec:7}), 
proving the above-suggested equivalence may pave the way to one solution of
quantum gravity.  But insofar as this `no boundary condition proposal' is
still somewhat  speculative (work in progress), in the remainder of this 
paper I will adhere to
the interpretation of the action given in the previous paragraph, namely that
$h_{\mu\nu}(\hat{e})$ is a fixed background metric, and only $U$ is dynamical
on $\partial M$.

\section{The Boundary Theory Phase Space}
\label{sec:5}
\noindent
In the previous sections I have attempted to carefully describe my motivations
for studying the action $I^{(1)}[e]$, and in the remainder of the paper I
will quickly outline some of the main results concerning its most interesting
aspect, namely the boundary theory it contains.

Up to an unimportant sign, which depends on the signatures of the spacetime
and its boundary, the symplectic structure of the boundary theory has a
coadjoint orbit form given by
\begin{equation}
\label{eqn:symplectic_structure}
\omega=\pm{1\over 2\kappa}\int_{\partial\Sigma}\epsilon_{\partial\Sigma}\,
{\textstyle\rm Tr}\, \{ \hat{T}(U^{-1}\delta U)(U^{-1}\delta U)\}\, .
\end{equation}
Here $\partial\Sigma$ is a $(d-2)$-dimensional 
Cauchy surface
in $\partial M$, which can be thought of as the intersection of $\Sigma$, 
a $(d-1)$-dimensional Cauchy surface in $M$, with the boundary $\partial M$.
Using (\ref{eqn:Euler-Lagrange_equations}) it is easy to show that $\omega$
is invariant under deformations of $\partial\Sigma$ connected to the 
identity, which is to say that this is a covariant description of the
boundary theory phase space.  We will focus here on the Euclidean
black hole (with bifurcation $(d-2)$-sphere excised---see Section~\ref{sec:3}),
which
is topologically an annulus cross $S^{d-2}$.  Choosing a constant
(Euclidean) time surface $\Sigma=[r_i ,r_o ]\times S^{d-2}$, we have
$\partial\Sigma=S_{i}^{d-2}\cup S_{o}^{d-2}$, 
where $S_{o}^{d-2}$ is a large
sphere at infinity and $S_{i}^{d-2}$ is a sphere 
cross section of the boundary of the
thickened bifurcation sphere.  At any instant of time, $U$ is a map from
$\partial\Sigma$ into the rotation group, $G$.  Under time evolution this
map changes, and its history is the $G$-valued function $U$ on $\partial M$.
For example, in the $d=3$ case, $U$ is 
(two copies of) a time dependent map from $S^1$
($S_{i}^{1}$ and $S_{o}^{1}$) into $SO(3)$, also known as the loop group of
$SO(3)$.  $\epsilon_{\partial\Sigma}$ is the volume form induced on 
$\partial\Sigma$, and depends on `macrostate' information encoded in
$h_{\mu\nu}(\hat{e})$.  And finally, $\hat{T}$ is a fixed element of the
Lie algebra, $g$, which is (or rather $\mp\epsilon_{\partial\Sigma}\hat{T}$ 
is) the pullback of $\hat{E}$ to $\partial\Sigma$.  The heart of the boundary
theory is the orbit, $U\hat{T}U^{-1}$, of $\hat{T}$ under the action of $U$.

Now let $H$ denote the isotropy subgroup of $\hat{T}$, i.e., the elements
$V$ of $G$ satisfying $V\hat{T}V^{-1}=\hat{T}$.  Introducing the coset
decomposition
\begin{equation}
\label{eqn:coset_decompostion}
U=\tilde{U}V\, ; \;\;\; U\in G\, , \;
                        V\in H\, , \;
                \tilde{U}\in G/H
\end{equation}
the symplectic structure in (\ref{eqn:symplectic_structure}) reduces to
\begin{equation}
\label{eqn:reduced_symplectic_structure}
\omega=\pm{1\over 2\kappa}\int_{\partial\Sigma}\epsilon_{\partial\Sigma}\,
{\textstyle\rm Tr}\, \{ \hat{T}(\tilde{U}^{-1}\delta \tilde{U})
(\tilde{U}^{-1}\delta \tilde{U})\}\, ,
\end{equation}
which means that the reduced (or physical) phase space consists of the 
set of all maps
$\tilde{U}$ from $\partial\Sigma$ into the coset space $G/H$.\footnote{When
$U$ is a rotation which preserves the normal,
i.e., $e_{\perp}=\hat{e}_{\perp}$,
the boundary action is independent of the remaining free rotation angle(s).
This symmetry reduces the physical phase space further, but only by the
removal of a set of points of measure zero.  Since our goal is to evaluate
the degeneracy of microstates, which at the semiclassical level corresponds to
calculating a volume in the physical phase space, this subtlety should not
significantly affect our results, and I will ignore it here.}  The image of
$\partial\Sigma$ in $G/H$ determined by a given $\tilde{U}$ 
can be thought of as the classical state of the
system at a given instant of time, and this
image evolves with time.  For the Euclidean black hole
the maps we are dealing with are those 
from $S^1$ into $SO(3)/SO(2)\cong S^2$ ($d=3$ case),
and $S^2$ into $SO(4)/(SO(2)\times SO(2))\cong S^2 \times S^2$ ($d=4$ case).
In general, $G/H=SO(d)/(SO(2)\times SO(d-2))$, an oriented Grassmann
manifold, with
corresponding changes from compact to noncompact groups when the spacetime
or boundary have non-Euclidean signature.  The dimension of $G/H$ is 
$2(d-2)$, twice the dimension of $\partial\Sigma$, and our phase space
(which is of course infinite-dimensional) can be thought of as 
even-dimensional for all $d$.

What does this symmetry $H$ correspond to physically?  Let the gauge-fixed
vector frame $\hat{e}_a$ consist of: $\hat{e}_\perp$, normal to $\partial M$;
$\hat{e}_\top$, tangent to $\partial M$ and normal 
to $\partial\Sigma$---a time flow unit vector; 
and the remaining $(d-2)$ unit vectors,
$\hat{e}_\alpha$, tangent
to $\partial\Sigma$.  Then the $SO(2)$ component of $H$ corresponds to
rotations involving just $\hat{e}_\perp$ and $\hat{e}_\top$; the $SO(d-2)$
component to rotations involving just the $\hat{e}_\alpha$.  This means the
boundary theory considers as physical only an observer's orientation 
{\em relative to} $\partial\Sigma$, for instance the bifurcation sphere, or
the large sphere at infinity.

The maps from $\partial\Sigma$ into $G/H$ have a semi-direct product type
of structure consisting of reparametrizations of the image and deformations
of the image normal to itself (normal with respect to the natural metric
on the Grassmann manifold).  Applied in the simplest case, the $(d=3)$ 
Euclidean BTZ black hole~\cite{BTZ}, it turns out that the components
of the energy-momentum tensor $T^{\mu\nu}$ in 
(\ref{eqn:energy-momentum_tensor}) canonically generate both reparametrizations
and deformations, in certain combinations.  Their Poisson algebra has a
Virasoro algebra piece (presumably corresponding to the reparametrizations),
and an additional rather complicated piece depending on the acceleration,
or extrinsic curvature of the image---in this 
($d=3$) case a closed curve $S^1$ in
$S^2$ (presumably corresponding to the deformations).  The Hamiltonian, equal
to the invariant length of the image curve, canonically generates
reparametrizations (only), the rate of reparametrization depending on the
extrinsic curvature of the image curve at the point in question.  
The simplest solutions to (\ref{eqn:Euler-Lagrange_equations})
are curves of constant extrinsic curvature (e.g., latitudes of the sphere),
this constant being restricted to an infinite but discrete set of values
(discrete because the solutions must be periodic in Euclidean time, with
period equal to the inverse temperature of the black hole).  These solutions
have the form of free fields propagating on $\partial M$.  The case of 
non-constant extrinsic curvature,
which can be solved only implicitly, describes an apparently much larger sector
of the solution space in which the solutions generically exhibit a 
``shock discontinuity'' phenomenon of the type discussed in 
Ref.~\cite{Cour} (and familiar to those studying fluid dynamics).

Returning to the general case we make the following
interesting observation: The $U$ boundary degrees of freedom, which I am
suggesting are ultimately responsible for black hole entropy, are
intimately associated with the group {\em Diff}${}\,(\partial\Sigma)$, in
particular the diffeomorphism group of the bifurcation $(d-2)$-sphere.
This type of result was speculated by Carlip in the conclusion section of
Ref.~\cite{Carl3}, and it is satisfying to see a concrete realization
of this idea.  And notice that the diffeomorphism symmetry of the
action $I^{(1)}[e]$ is not broken in order to achieve this.
Nor do diffeomorphisms of the boundary play any role as gravitational
boundary degrees of freedom.  (I mention this because there are interesting
discussions to the contrary which suggest that diffeomorphisms of the boundary,
perhaps even those which do not preserve the boundary, 
might represent gravitational boundary degrees of
freedom~\cite{Bala,Carl3,Bana2}.  
While this might be true, it is not obvious to me how
the results here could be connected with that idea, except perhaps through
the `no boundary condition proposal' discussed in Section~\ref{sec:4}.)
Anticipating
some possible confusion on this point, let me restate that  
the action is invariant under diffeomorphisms of $e$ 
(read: {\em simultaneous} diffeomorphisms of 
$U$ and $\hat{e}$) which preserve $\partial M$.  
On the other hand, while it
is true that a reparametrization of the image of $\partial\Sigma$ in
$G/H$ is related to a diffeomorphism of the scalar field $U$ on $\partial M$,
since $\hat{e}$ is not simultaneously undergoing the same diffeomorphism
transformation the action is in general not invariant.  In other words,
reparametrizations of the image are not, in general, (diffeomorphism) 
symmetries of the action, but in fact
are physical degrees of freedom.

\section{Generalized Ka\v{c}-Moody Algebras and Virasoro Operators}
\label{sec:6}
\noindent
$\partial\Sigma$ may consist of a number of disconnected components;
in an abuse of notation I will use $\partial\Sigma$ to denote any one
of these components, which are typically $(d-2)$-spheres.
We introduce a mode basis, $f_M$, $M$ a
collective index of integers, which satisfy the following relations:
\begin{eqnarray}
\int_{\partial\Sigma}\epsilon_{\partial\Sigma}\, \bar{f}_{M}f_{N} &=&
\delta_{MN}\, {\textstyle\rm Vol}\, (\partial\Sigma)\, , \;\;\;
{\textstyle\rm where} \;\;  
{\textstyle\rm Vol}\, (\partial\Sigma)=
\int_{\partial\Sigma}\epsilon_{\partial\Sigma}\, ,
\label{eqn:orthonormality} \\
f_M f_N &=& \sum_P \bar{C}_{MNP}\, f_P \, .
\label{eqn:mode_structure_constants}
\end{eqnarray}
Here $\bar{f}_{M}$ denotes the complex conjugate of $f_{M}$ in case the mode
basis is complex.
The $C_{MNP}$ are mode structure constants analogous to the familiar
Clebsch-Gordon coefficients used in the quantum mechanical treatment of
angular momentum,  which are calculated
using (\ref{eqn:orthonormality}).\footnote{Notice that the $C_{MNP}$
might encode information about the `macrostate' not already contained in
${\textstyle\rm Vol}\, (\partial\Sigma)$, such as information about a
conformal `weight factor' in the measure, or the Teichm\"{u}ller
parameters of the metric on $\partial\Sigma$, and that this information will
be reflected in the quantization.  I have not yet investigated this 
possibility in any detail.}  
Recall that the heart of the boundary theory is the orbit of $\hat{T}$ under
the action of $U$: there is a Lie algebra-valued current,
$J$, proportional to $U\hat{T}U^{-1}$ ($= \tilde{U}\hat{T}\tilde{U}^{-1}$).
Defining the modes of this current as
\begin{equation}
\label{eqn:current_modes}
J_M = \mp {1\over\kappa} \int_{\partial\Sigma} \epsilon_{\partial\Sigma}\,
\bar{f}_M \,
(\tilde{U}\hat{T}\tilde{U}^{-1})\, ,
\end{equation}
it can be shown that the symplectic structure given in 
(\ref{eqn:reduced_symplectic_structure}) is equivalent to the Poisson bracket
algebra
\begin{equation}
\label{eqn:general_kac-moody}
\{ J^{A}_{M},J^{B}_{N} \} = - f^{AB}_{\;\;\;\;C} \sum_P C_{MNP} \, 
J^{C}_{P} \, ,
\end{equation}
with the currents being subject to the quadratic constraints
\begin{equation}
\label{eqn:general_constraints}
L_P = 
\left( {{{\textstyle\rm Vol}\, (\partial\Sigma)}\over{\kappa}} \right)^2 \, 
\delta_{PP_0}\, , \;\;\;
{\textstyle\rm with} \;\; L_P =
{1\over{\hat{N}}} \sum_{M,N} C_{PMN}\, N_{AB}\,
\bar{J}^{A}_{M}J^{B}_{N} \, .
\end{equation}
My Lie algebra basis notational conventions are 
$[ T_A , T_B ] = f_{AB}^{\;\;\;\;\;\,C} T_C$ and 
${\textstyle\rm Tr}\, ( T_A T_B ) = - 2 N_{AB}$, where $N_{AB}$ and its
inverse are used to raise and lower the Lie algebra indices.
These indices run from 1 to $d(d-1)/2$.
$\hat{N}$ is defined to be $-{\textstyle\rm Tr}\,( \hat{T}\hat{T})/2$.
$P_0$ is the unique value of the collective index $P$ such that 
$f_{P_0}=1$, the constant mode.  With these conventions we have 
$C_{P_{0}MN} = \delta_{MN}$, so that $L_{P_0}$ is real and positive
definite (at least in the Euclidean case).

The constraints in (\ref{eqn:general_constraints}) follow from the fact that
${\textstyle\rm Tr}\,(JJ)$ equals a constant: the $J^{A}_{M}$ comprise an
overcomplete set of phase space coordinates.  As a simple analogy, consider
a two-sphere of radius $R$ (with spherical coordinates $(\theta,\phi)$)
as a phase space with canonical coordinates $p=-R\cos\theta$,
$q=\phi$, such that $\{ p,q\} =1$.  It is often more convenient to use
instead the overcomplete set of phase space coordinates 
$J^{1}=R\sin\theta\cos\phi$, $J^{2}=R\sin\theta\sin\phi$, $J^{3}=R\cos\theta$
satisfying an angular momentum type of algebra $[J^1 ,J^2 ]=J^3$, 
etc.\ (the analogue of (\ref{eqn:general_kac-moody})), 
and subject to the constraint $|J|^{2}=R^{2}$
(the analogue of (\ref{eqn:general_constraints})).  
Notice that, like its analogue $R$, 
${\textstyle\rm Vol}\, (\partial\Sigma)/\kappa$ in
(\ref{eqn:general_constraints}) sets the
{\it scale} of the phase space.

These constraints are
similar to those encountered in string theory, where the $L_{P_0}$ constraint
here is analogous to what is known there as the mass-shell condition
\cite{Gree}.  In the case $d=4$ let $M_s$ be the string mass and let
$\alpha^\prime$ denote what is known as the universal Regge slope
parameter (inversely proportional to the string tension).  Then the
role of $\sqrt{\alpha^\prime}M_s$ in string theory is played here by
${\textstyle\rm Vol}\, (\partial\Sigma)/\kappa$; the latter can be taken to
be proportional to the area of the bifurcation sphere in Planck
units, i.e., $GM^{2}_{bh}$, where $G$ is Newton's 
constant and $M_{bh}$ is the black hole mass.  As argued in 
Ref.~\cite{Horo2}
the string coupling should be chosen such that 
$\sqrt{\alpha^\prime}/G\sim M_{bh}$, which means $M_{s}\sim M_{bh}$.
This identification of masses suggests that the black hole can be viewed
as an excited string state~\cite{Horo}. 
Now, the mass-shell condition determines $M_s$ in terms
of the internal vibrational modes of the string.  Based on the preceding
discussion it thus seems reasonable to suggest that the $L_{P_0}$ constraint,
in some similar way, associates the mass of the black hole (or the
area of its bifurcation sphere) with the $U$ vibrational modes on 
its horizon (more precisely,
the boundary of its thickened bifurcation sphere).
Additional hints that the analysis here might have hidden connections
with string theory are discussed in Section~\ref{sec:7}.

Let us see what (\ref{eqn:general_kac-moody}) and 
(\ref{eqn:general_constraints}) look like 
when $d=3$.  In this case take $\partial\Sigma=S^1$; then 
$f_{M}=f_{m}(\phi)=e^{im\phi}$, $m$ an integer, $C_{mnp}=\delta_{m+n,p}$, and
(\ref{eqn:general_kac-moody}) reduces to the (classical) Ka\v{c}-Moody
algebra
\begin{equation}
\label{eqn:d=3_kac-moody}
\{ J^{A}_{m},J^{B}_{n} \} = - f^{AB}_{\;\;\;\;C}
J^{C}_{m+n} \, ,
\end{equation}
and the constraints in (\ref{eqn:general_constraints}) take the form
\begin{equation}
\label{eqn:d=3_constraints}
L_0 = 
\left( {{{\textstyle\rm Vol}\, (\partial\Sigma)}\over{\kappa}} \right)^2 
\, ; \;\;\;
L_m = 0 \, , \; m>0
\, ; \;\;\;
L_m =
{1\over{\hat{N}}} \sum_{n} N_{AB}\,
\bar{J}^{A}_{n-m}J^{B}_{n} \, .
\end{equation}
(The constraints $L_{m}=0$, $m<0$ are accounted for by the
reality condition $\bar{L}_m =L_{-m}$, which follows from 
$\bar{f}_m = f_{-m}$.  Just as in the Gupta-Bleuler treatment of
electrodynamics, this observation is necessary to make sense of such
constraints at the quantum level---see, e.g., Section~2.2 of
Ref.~\cite{Gree}.)  The $L_m$ are known as Virasoro
operators (or rather, their classical counterparts), 
and are closely connected with certain components of the 
energy-momentum tensor $T^{\mu\nu}$ in (\ref{eqn:energy-momentum_tensor}).
Much is known about the Ka\v{c}-Moody algebra and its associated Virasoro 
algebra (an excellent review can be found in 
Ref.~\cite{Godd}), but the point I would like to emphasize here 
is that the
$L_m$ are generators of the group {\em Diff}${}\,(S^{1})$.  
It seems plausible
that (\ref{eqn:general_kac-moody}) and 
(\ref{eqn:general_constraints})---higher-dimensional generalizations of
(\ref{eqn:d=3_kac-moody}) and 
(\ref{eqn:d=3_constraints})---are in like manner
associated with the group {\em Diff}${}\,(\partial\Sigma)$, in particular
{\em Diff}${}\,(S^{d-2})$, of which much less is known.  
Of greatest interest
is the case $d=4$.  Starting with 
the fact that $so(4)\cong so(3)\oplus so(3)$, it can
be shown that (\ref{eqn:general_kac-moody}) and 
(\ref{eqn:general_constraints}) reduce to simply two commuting copies of
the $d=3$ case, (\ref{eqn:d=3_kac-moody}) and 
(\ref{eqn:d=3_constraints}), except that we must use the mode structure
constants $C_{MNP}$ of the spherical harmonics in place of the simpler
$C_{mnp}=\delta_{m+n,p}$.  So, at least from this
point of view, 
understanding {\em Diff}${}\,(S^{2})$, and
eventually the microstates responsible for the entropy of the 
four-dimensional black hole (in the approach considered here), 
does not seem that far out of reach.

\section{Quantization, Microstates, and Black Hole Entropy}
\label{sec:7}

Before discussing quantization of the boundary theory let us consider the
following.  Boltzmann's formula tells us that the entropy, $S$, of a
physical system is given by $\ln\Gamma$, where $\Gamma$ is the number of
microstates compatible with the macrostate the system is in.  At the
semiclassical level $\Gamma$ is identified with (in a way I will not
attempt to make precise here) the volume of the phase space (or perhaps
some subset of it).  In Section~4 we found that the (reduced) phase space
is the set of all maps from $\partial\Sigma$ (a constant Euclidean time
slice of $\partial M$) into the Grassmann manifold
${\textstyle\rm Gr}\, (2,d-2)=SO(d)/(SO(2)\times SO(d-2))$.  For a Euclidean
black hole with $M=R^2 \times S^{d-2}$ we have $\partial\Sigma=S^{d-2}$.  (As
argued in Section~2 we need only consider the inner boundary, i.e.,
we can take $\partial\Sigma$ to be a constant Euclidean time slice of just
the boundary of the thickened bifurcation sphere.)  Now clearly our whole
analysis would go through unchanged if we instead considered
$M=R^2 \times {\cal S}^{d-2}$, where ${\cal S}^{d-2}$ is any compact
$(d-2)$-dimensional manifold without boundary, in which case
$\partial\Sigma={\cal S}^{d-2}$.  So our phase space would consist of the
set of all maps from ${\cal S}^{d-2}$ into ${\textstyle\rm Gr}\, (2,d-2)$.
To calculate the volume of this space (i.e., to determine $\Gamma$) would
require a measure and careful regularization, which I will not attempt
to do here.  Instead I will give just a rough sketch of how one might proceed.

In place of the volume of the phase space let us think of $\Gamma$ as the
number of ways of embedding ${\cal S}^{d-2}$ into 
${\textstyle\rm Gr}\, (2,d-2)$, treated as a problem in combinatorics:
Partition ${\cal S}^{d-2}$ into a number, $\alpha$, of cells, and 
${\textstyle\rm Gr}\, (2,d-2)$ into $\beta$ cells.  There is a natural
choice for $\alpha$.  Let $A$ denote the `area' (or volume) of 
${\cal S}^{d-2}$ which, in the limit of infinitesimal thickening
(of the excised bifurcation $(d-2)$-surface), is equal to the area of the
bifurcation $(d-2)$-surface.  
It is natural to take $\alpha=A/\kappa$, where $\kappa$ is 
(a numerical factor times)
the Planck area.  Since each cell of ${\cal S}^{d-2}$ can be embedded
independently into any one of the cells of ${\textstyle\rm Gr}\, (2,d-2)$
(ignoring smoothness considerations),
we have $\Gamma=\beta^\alpha$.  So we find that $\ln\Gamma$ is
proportional to $A/\kappa$ (in any dimension, $d$), the answer we want.
Admittedly this is a simple-minded argument.  Nevertheless, although a 
more sophisticated argument might determine the proportionality constant,
I find it hard to imagine how it could give a qualitatively different
result (e.g., $\ln\Gamma$ depending nonlinearly on $A/\kappa$, perhaps
even in a way involving the dimension, $d$)---this point will be important
later.

But this is not the end of the story.  I think the intriguing part of it is
this: It is well known that the homotopy classes of maps from 
${\cal S}^{d-2}$ into a Grassmann manifold are intimately connected with
the theory of characteristic classes, in particular the Euler number,
$\chi$, of ${\cal S}^{d-2}$.  (A relevant specific example can be found
in Ref.~\cite{Cher}.)  So it seems quite plausible that a careful
semiclassical calculation of $\ln\Gamma$ will produce not only a factor
proportional to $A/\kappa$, but will also involve $\chi$.  Finally,
observing that $\chi$ is also the Euler number of $M$ 
($=R^{2}\times {\cal S}^{d-2}$ in our case), the above considerations
are quite suggestive of the generalized entropy formula $S=(\chi/8)A$
given in (\ref{eqn:generalized_entropy})~\cite{Libe}.

Now let us move from these semiclassical considerations to a quantization
proper of the boundary theory.  Equation
(\ref{eqn:general_kac-moody}) displays the fundamental Poisson bracket
relations (analogous to $\{ p,q\}=1$) that on quantization will yield an
infinite-dimensional Hilbert space of states (the microstates).  
A finite-dimensional
subspace of {\em physical} microstates is obtained by imposing the
constraints (\ref{eqn:general_constraints}) as quantum operators acting on
these microstates.  The dimension of this space is now the $\Gamma$
to be used in Boltzmann's formula $S=\ln\Gamma$, and can be thought
of as essentially the degeneracy of microstates satisfying the $L_{P_0}$
`mass-shell constraint.'

Unfortunately, I do not yet know how to carry out this quantization
(and counting) program for cases $d>3$.  Backtracking somewhat, the
quantization program can be described as the task of finding
irreducible unitary highest weight representations of the Lie algebra
corresponding to the group of all maps from $\partial\Sigma$ into $G$ 
($=SO(d)$ in the Euclidean case), subject to certain constraints
(the quantum analogue of the classical reduction from $U$
to $\tilde{U}$---see (\ref{eqn:symplectic_structure}) and 
(\ref{eqn:reduced_symplectic_structure})).  Finding such representations
(in numerous contexts)
is currently an active area of 
research (see, e.g., Refs.~\cite{Fadd,Ferr,Cede,Mick1,Lars,Mick2,West}). 
Interest originated in the
study of anomalous gauge theories with chiral fermions~\cite{Fadd}, 
and more recently
has seen applications in $p$-branes and the study of non-perturbative
effects in string theory~\cite{Ferr,Cede}.
(Ref.~\cite{Ferr} contains at least a partial list of the many important
applications of this branch of representation theory to physics.)
Unfortunately, at present there remain many
unresolved issues.  For instance, it is often difficult to prove the
unitarity of attempted representations~\cite{Cede,Mick1}; 
and instead of a simple c-number central
charge (see below) 
one encounters a number of possible operator-valued Schwinger
terms depending in a complicated way on additional external 
fields~\cite{Ferr,Cede,Mick1,Mick2,West}.  
We will see below, in the $d=3$ case, that a certain c-number central charge
plays a pivotal role in the physics of entropy.  I expect that the 
operator-valued Schwinger terms occurring in higher dimensions will play
an equally important role, but it is precisely these terms which are
difficult to determine.
Thus I will not attempt
to pursue a quantization of the $d>3$ cases here.

Fortunately the $d=3$ case involves simply a loop group; much is known
about quantizing the Euclidean version of equations 
(\ref{eqn:d=3_kac-moody}) and 
(\ref{eqn:d=3_constraints}) (see, e.g., Ref.~\cite{Godd}).
The (quantum) Ka\v{c}-Moody algebra, obtained by replacing the Poisson bracket
with $i$ times the commutator bracket, is
\begin{equation}
\label{eqn:quantum_d=3_kac-moody}
[J^{A}_{m},J^{B}_{n}] = if^{AB}_{\;\;\;\;C} J^{C}_{m+n}
+km\delta^{AB}\delta_{m+n,0} \, .
\end{equation}
The additional term on the far right is a central extension.  The c-number
central charge, $k$, is of purely quantum mechanical origin---I emphasize 
that our classical analysis gives us no information on the value of this
number.
Now given (\ref{eqn:quantum_d=3_kac-moody}) and the quantum version of
(\ref{eqn:d=3_constraints}) an asymptotic formula for $\Gamma$ 
is available~\cite{Kac}:
\begin{equation}
\label{eqn:main_result}
\ln\Gamma\sim F(k)\, 
{{{\textstyle\rm Vol}\, (\partial\Sigma)}\over\kappa}
\end{equation}
in the limit 
${\textstyle\rm Vol}\, (\partial\Sigma)/\kappa\rightarrow\infty$.  Again,
strictly speaking, the right hand side of (\ref{eqn:main_result}) is a sum
over the disconnected components of $\partial\Sigma$.  But
as argued in Section~\ref{sec:3}, and mentioned again in the semiclassical
discussion above, we need only consider the component of $\partial\Sigma$
which is a constant Euclidean time slice of the boundary of the thickened
bifurcation sphere.  So in the limit of infinitesimal thickening,
${\textstyle\rm Vol}\, (\partial\Sigma)=A$, the `area' (in this case
length) of the bifurcation one-sphere of the Euclidean BTZ black hole.
Thus, with $\kappa$ a Planck scale unit of area, $A/\kappa$ is
large for any black hole much larger than the Planck scale,
and our analysis says that in this case the physical
entropy is given by
\begin{equation}
\label{eqn:main_result_entropy}
S=F(k)\,{A\over\kappa}\, .
\end{equation}
The function $F$ of the central charge $k$ can, in principle, 
be worked out using results in Ref.~\cite{Kac},
but for the purposes of the remainder of this section we will not need to
know its precise form.

At this point I can state three main observations
emerging from the analysis.  First, the boundary theory 
contained in $I^{(1)}[e]$ leads to physical microstates 
living on the boundary of a thickened bifurcation sphere which are 
{\em sufficiently rich in number} to 
provide a statistical mechanical account of the
entropy of the $d=3$ Euclidean black hole: our answer for the entropy is
a (variable) numerical factor times the horizon area.  
Let us pause to ask whether we can expect a similar result to be obtained
in any dimension, $d$.  
In other words, will this approach---at the quantum level---give
an entropy proportional to the area in any dimension?
In the $d=3$ case the factor $A/\kappa$ in (\ref{eqn:main_result_entropy})
comes from the square root of the constrained value of $L_0$ in the
mass-shell constraint in (\ref{eqn:d=3_constraints}).  The higher
dimensional analogue of (\ref{eqn:d=3_constraints}) is
(\ref{eqn:general_constraints}), which has exactly the same quadratic
dependence on ${\textstyle\rm Vol}\, (\partial\Sigma)/\kappa$.  So the
question is, will the number-theoretic counting arguments for cases
$d>3$ give a degeneracy of microstates depending on the {\it square
root} of $({\textstyle\rm Vol}\, (\partial\Sigma)/\kappa)^2$, as it does
in the $d=3$ case?  Unfortunately the answer is not at all obvious
because the counting arguments depend
intimately on the details of quantization.\footnote{I
am indebted to Steven Carlip for pointing this out to me.}  And insofar
as the quantization of the boundary theory for cases $d>3$ is fraught
with technical challenges (some of which were mentioned above), I do
not yet know the answer.  However, the rough semiclassical argument
sketched at the beginning of this section provides hope that this question
will be answered in the affirmative.

The second main observation concerns the numerical factor multiplying
the area (more precisely, $A/\kappa$) in the entropy results.  
The rough semiclassical argument
suggests that this factor will be related to the Euler number, $\chi$,
and hence is of topological origin.  
On the other hand, the $d=3$ quantum calculation
resulting in (\ref{eqn:main_result_entropy}) shows that this factor
($F(k)$) depends on a central charge; since this central charge
is not present in the classical or semiclassical analyses this result
emphasizes how deeply quantum mechanical the phenomenon of black hole
entropy really is.  (Observe that the $\hbar$ in the Planck-scale factor
$\kappa$ also, of course, shows that black hole entropy is a quantum
phenomenon, but it is independent of the {\it details} of quantization,
being present already at the semiclassical level.)
This is highly intriguing on several accounts.  At the
most basic level it represents a concrete realization of a suggestion
by Carlip and Teitelboim that the numerical factor relating the
entropy of a Euclidean black hole to the area of its bifurcation sphere
might have its origin in microstates living on 
the boundary of a thickened bifurcation
sphere~\cite{Carl2}.  
$F(k)$ is precisely this numerical factor.
At a deeper level this observation hints at a possible connection
with recent results in quantum geometry and black hole entropy
achieved in the framework of non-perturbative 
canonical quantum gravity~\cite{Asht1,Asht2}.
The spectrum of the area operator in that approach
is known to be discrete, and have a minimum eigenvalue of zero, with
the next eigenvalue---the ``area-gap''---depending on the topology
of the surface in question.  In this sense ``quantum geometry `knows'
about topology''~\cite{Asht2}.  This is somewhat analogous to what appears to
be happening here.  If the semiclassical argument presented here is
correct, then the purely quantum factor $F(k)$ must be proportional to
the Euler number (more properly speaking---in this one-dimensional 
case---some topological invariant
applicable for odd-dimensional surfaces): the quantum
microstates on the thickened bifurcation surface `know' about the
topology of the bifurcation surface.  More generally, are `quantum anomalies'
of a diffeomorphism algebra associated with topology?

Consider also the following.  In the 
non-perturbative canonical quantum gravity approach to black hole microstates
as described in Ref.~\cite{Asht2} it is pointed out that, strictly
speaking, at the classical level there are no independent boundary degrees 
of freedom; it is only at the quantum level that independent boundary 
degrees of freedom arise, which then account for the black hole entropy.
This is apparently in marked contrast to what happens here, where
there {\it are} independent boundary degrees of freedom beginning at the
classical level.  But is this apparent contrast really so?  
Although we begin with classical degrees of freedom $U$, I emphasize
again that the central charge in (\ref{eqn:quantum_d=3_kac-moody}), and
hence $F(k)$ in the entropy formula (\ref{eqn:main_result_entropy}), is
of purely quantum mechanical origin---it has no classical analogue: again,
black hole entropy is a deeply quantum phenomenon.

The third main observation concerns
the {\it nature} of the result in (\ref{eqn:main_result_entropy}).  
It is well known, by a number of arguments independent of the analysis
presented here, that the entropy
of a black hole is given by
\begin{equation}
\label{eqn:generic_entropy}
S=\sigma\,{A\over\kappa}\, ,
\end{equation}
where $\sigma$ is a known numerical factor (e.g., $2\pi$ for nonextremal
black holes in four dimensions~\cite{Beke2}).
Our analysis does not yield $\sigma$,
but rather by {\em demanding} that (\ref{eqn:main_result_entropy})
reproduce the correct result (\ref{eqn:generic_entropy}), a unique numerical
value for the central charge $k$ is thus determined.  I believe that this
is significant.
It is known that the irreducible unitary highest weight representations of
the Virasoro algebra are determined completely by two numbers.  
The first is the eigenvalue of $L_0$ (analogous to the eigenvalue of $|J|^2$ 
in a quantum mechanical treatment of angular momentum), 
and the second is the value of $k$~\cite{Godd}.  
In our case, the first is proportional to $(A/\kappa)^2$, 
and the second is determined
by equating (\ref{eqn:main_result_entropy}) with (\ref{eqn:generic_entropy}).
In this way a particular representation of the 
Virasoro algebra is determined by the analysis.
Furthermore, regarding
representations of the Ka\v{c}-Moody algebra, 
it is known that the vacuum states must form a
representation of $g$, and I suspect that certain vacuum solutions of
the boundary equations will play a role here, but this remains to be
investigated.  

The question of how to fix $k$ has an analogue in the non-perturbative
canonical quantum gravity approach to black hole microstates \cite{Asht1}.
As mentioned above in the Introduction, there the entropy of a large
non-rotating black hole is found to be proportional to $A/\kappa$ divided
by the Immirzi parameter, $\gamma$.  An appropriate choice for $\gamma$
then yields the Bekenstein-Hawking entropy---the same argument used in the
previous paragraph to determine $k$.  It is worth noting an additional
parallel between these two variable parameters: on the one hand, different
values of $k$ correspond to unitarily inequivalent representations of 
the Virasoro algebra (and different spectra of the Virasoro operators);
on the other hand, different
values of $\gamma$ correspond to unitarily inequivalent representations of 
the canonical commutation relations (and a different spectrum of the 
area operator in loop quantum gravity)~\cite{Immi,Rove}.
In Ref.~\cite{Asht1} it is remarked that the ambiguity in choosing
$\gamma$ can be resolved only with the help of additional input
(such as demanding agreement with the Bekenstein-Hawking entropy 
formula).\footnote{In some respects this argument might seem unsatisfactory:
shouldn't a statistical mechanical analysis uniquely determine the entropy
without recourse to thermodynamics?  It is worth pointing out that an
`internal' mechanism might be available in our case to do just so.  Work in
progress indicates that the function $F(k)$ is peaked at a unique value
of $k$.  If correct, this would provide a very satisfying physical
argument to fix $k$; namely, choose $k$ such that the entropy is 
maximized.}

Another point regarding the importance of the value of the central charge
lies in a result known as the 
``quantum equivalence theorem''~\cite{Godd}. 
This theorem may have interesting consequences for the analysis here.
As an illustration of this theorem, in a 
phenomenon he called non-Abelian bosonization, Witten argued that the
``level 1'' (read: a certain value of $k$) $SO(N)$ WZW theory
is ``quantum equivalent'' to a theory of 
$N$ massless free fermions~\cite{Witt}.
This is surprising, given that the two theories look quite different, and
note that the result depends crucially on the value of $k$.
Now consider that, on the one hand, Carlip's analysis~\cite{Carl} 
involves a WZW boundary theory; on the other hand, it is
not inconceivable that our boundary action, 
being of the form $U^{-1}dU$, with $U$ an orthogonal matrix, 
could be cast into a Dirac form involving spinors.
[Further support for this possibility comes from work by
Baez {\it et al} on the quantum gravity Hamiltonian for manifolds with
boundary, which hints at the existence of a boundary theory of Weyl spinors
\cite{Baez}.  This work is in the context of the loop variables approach to
quantum gravity, and given the receptiveness of the Goldberg action
(and hence $I^{(1)}[e]$ in four dimensions) to an Ashtekar variables 
formulation~\cite{Gold,Lau1,Lau2}, it is tempting to seek a connection
between the work in~\cite{Baez} and what is done here.]
So it might be possible to demonstrate ``quantum equivalence'' between the two
approaches using ideas analogous to Witten's non-Abelian bosonization.
Such a result would presumably hinge on the precise numerical value of $k$.
In any case, although the $d=3$ boundary theory here and the one in 
Ref.~\cite{Carl} look quite different, they both purport to describe
the same physics, and certainly the mathematics used in the final stages
of counting microstates is strikingly similar: these are strong indications
that there exists some hidden connection between the two 
descriptions.\footnote{One additional remark can be made.  The current
algebra in Carlip's analysis~\cite{Carl} is an $so(2,1)\oplus so(2,1)$
version of (\ref{eqn:quantum_d=3_kac-moody}), with a central charge I
will denote as $\tilde{k}$.  Unlike my $k$, $\tilde{k}$ depends on the
cosmological constant, $\Lambda$.
Carlip argues that in the semiclassical regime $\Lambda$ is small, making
$\tilde{k}$ large, which has the effect of `Abelianizing' his
current algebra, and thus simplifying the task of counting microstates.
(Remark: although Carlip is making a large $\tilde{k}$ approximation his
calculation does depend on the value of $\tilde{k}$, and in such a way that, 
of course, the final result for the entropy is independent of $\Lambda$.)  
Notice that
this would seem to suggest that the non-Abelian nature of
the current algebra is not important to the final result, in marked contrast
to the analysis here, in which a large $k$ approximation option does not
make any sense, and the non-Abelian nature of (\ref{eqn:quantum_d=3_kac-moody}) 
is crucial.  Apparently $k$ and $\tilde{k}$ play very different roles.
A hint towards understanding the relationship between Carlip's approach and
the one here comes from work by Ba\~{n}ados and
Gomberoff~\cite{Bana3} in which they argue that in fact the non-Abelian
nature of the current algebra {\em does} play an important role in
Carlip's analysis, except that this role is inadvertently hidden: if
one makes a large $\tilde{k}$ approximation at an earlier stage of
the calculation, rather than near the end,
one does not get the correct result for the entropy.  (I am indebted to
Jack Gegenberg for bringing Ref.~\cite{Bana3} to my attention, and 
to Steven Carlip for clarification of his analysis.)}

Finally, we have already seen several hints that the approach discussed
here might be related to the string theory approach to black hole microstates.
As another, perhaps more direct hint,
the discussion in Section~4.5 of Ref.~\cite{Godd} establishes the
rather surprising result that the Virasoro operators associated with a Lie
algebra $g$, and those associated with a subalgebra $h\subset g$, are
``quantum equivalent'' provided merely that the two algebras have the same 
central charge.  This suggests it might be possible to show---again, depending
crucially on the value of $k$---that the boundary theory here is ``quantum
equivalent'' to a string moving on the maximal torus of the rotation group,
$G$.  (In this regard, observe that the isotropy
subgroup, $H$, in the $d=3$ and 4 cases is also, in fact, the maximal
torus subgroup of $G$.)

Another question which I have not yet touched on concerns corrections to the
Bekenstein-Hawking entropy formula.  It seems that the proposed action
$I^{(1)}[e]$ provides a clean and simple way to formulate this question.
Let $\bar{I}^{(1)}[e]=I^{(1)}[e]-I^{(1)}[e_{\rm vac}]$ 
denote the regularized Euclidean action, where $e_{\rm vac}$ is a suitable
vacuum spacetime solution, as discussed in Section~\ref{sec:3}.  (Notice
that $e_{\rm vac}$ consists of $\hat{e}_{\rm vac}$ as well as $U_{\rm vac}$.)
The partition function is given by
\begin{equation}
\label{eqn:partition_function}
Z=\sum_{e}\,[de]\,\exp \bar{I}^{(1)}[e]\, .
\end{equation}
We then write the measure as a product, $[de]=[d\hat{e}]\,[dU]$, 
and use the split action given in
(\ref{eqn:split_action}), suitably regularized.  As a first approximation
to $Z$ we retain only the classical piece $\hat{e}_{\rm cl}$ 
in the sum over $\hat{e}$, 
corresponding to a classical black hole `macrostate' solution.
In this approximation the partition function reduces to
\begin{equation}
\label{eqn:approximate_partition_function}
Z\approx \exp(\bar{I}^{(1)}[\hat{e}_{\rm cl}])\,
\sum_{U}\,[dU]\,\exp\bar{I}_{B}[U;\hat{e}_{\rm cl}]\, .
\end{equation}
The exponential outside the sum is responsible for the Bekenstein-Hawking
entropy formula.  The functional measure $[dU]$ is based on
the finite-dimensional Haar measure of the frame rotation group, $G$; its
``functional'' aspect is associated with the  
manifold $\partial M$, in particular, the boundary of the thickened
bifurcation sphere, which is {\em closed and compact}.  The
integrand in $I_{B}[U;\hat{e}_{\rm cl}]$ is proportional
to ${\textstyle\rm Tr}\, (U^{-1}dU\wedge\hat{E}_{\rm cl})$ 
(see (\ref{eqn:boundary_action})), and thus it is plausible that the sum
on the right hand side of (\ref{eqn:approximate_partition_function})
involves a functional determinant of the form 
${\textstyle\rm Det}_{\partial M}(d_{\rm cl})$.
Here $d_{\rm cl}$ denotes a linear differential operator on $\partial M$
constructed out of the exterior derivative, $d$, and $\hat{E}_{\rm cl}$;
notice that this operator contains information about the `macrostate'
(such as the black hole mass and angular momentum).  
Also, the boundary action
$I_{B}[U;\hat{e}]$ is invariant
under $U\rightarrow U_{0}U$, $U_{0}$ a constant element
of $G$ on $\partial M$; this symmetry may play a role in the zero modes of
$d_{cl}$.  In any case, 
given its simple form, it seems likely that the entropy 
correction implied by
(\ref{eqn:approximate_partition_function}) can probably be worked out
{\em nonperturbatively}.  Now, as emphasized by Carlip~\cite{Carl3}, 
``any quantum mechanical statement about black holes is necessarily a 
statement about {\em conditional} probabilities.''  This
means the following.  Suppose we have a complete theory of quantum gravity,
and $|\,bh,\psi\rangle$ denotes an eigenstate in which $bh$ (a
certain classical black hole), and $\psi$ (a certain set of values for
all the other classical configuration space degrees of freedom of our
hypothetical theory), are sharply defined.  When we make a quantum
mechanical statement about black holes we are restricted to transition
amplitudes of the form $\langle\,bh,\psi^{\prime}\,|\,bh,\psi\rangle$.
It is precisely this restriction which is reflected in the approximation
leading to (\ref{eqn:approximate_partition_function}), where
$\hat{e}_{\rm cl}$ corresponds to $bh$, and $U$ to $\psi$.
To the extent that a formula for the entropy of a black hole makes sense
only within this restriction, (\ref{eqn:approximate_partition_function})
would appear to contain {\em all} of the relevant 
quantum gravitational physics.  The possibility
that (\ref{eqn:approximate_partition_function}) can be solved 
nonperturbatively is thus quite exciting. 

My final point, which I have already discussed in Section~2, 
but wish to properly
clarify here, regards the question of just where do the microstates
that account for black hole entropy live?  Although the action $I^{(1)}[e]$ 
is completely general, and in particular can be analyzed in a spacetime
of any signature, the main results presented here
have been in the context of Euclidean gravity, and so I will restrict my
comments to this case.  Assuming a Euclidean black hole topology 
$R^{2}\times S^{d-2}$, there is certainly an $S^{1}\times S^{d-2}$ 
boundary at infinity, and our analysis says that microstates necessarily live
here.  But their degeneracy (and hence the entropy associated with them)
is infinite.  However, as argued in Section~\ref{sec:3}, the suitable
vacuum spacetime corresponding to this black hole has precisely the same 
entropy, and when subtracted, yields a physical black hole entropy equal to
zero.  This means that the microstates living on the boundary at infinity
are not physical.  (Such a regularization procedure is exactly analogous
to what is usually done to calculate the Bekenstein-Hawking entropy 
starting from the action given in (\ref{eqn:gauge-fixed_action})
\cite{Brow1,Brow2,Brow3,Hawk,Gibb}.)  
But there is a subtlety here: as argued in 
Section~\ref{sec:3}, if we are to take $I^{(1)}[e]$ seriously as an action 
based on an orthonormal frame, rather than the metric, both the mathematics
and physics strongly suggest that we excise the bifurcation sphere from $M$
(ignoring for now any other set of points which might also have to be
excised).
This introduces a thickened bifurcation sphere whose 
$S^{1}\times S^{d-2}$ boundary (the inner boundary)
I will denote as $\partial M_{i}$.  This is
where the physically relevant microstates live.  But if we apply this
excision principle to the black hole we must be prepared to do the same
to the corresponding vacuum spacetime.  In the case $d=4$, for example,
this vacuum spacetime is topologically $R^{3}\times S^{1}$ and, arguing again
as in Section~\ref{sec:3}, we must excise the point set $\{0\}\times S^{1}$.
`Thickening' this point set introduces an $S^{2}\times S^{1}$ boundary
which we shall denote as $\partial M_{i}^{\rm vac}$.  Now choose any constant 
Euclidean time slice in $\partial M_{i}^{\rm vac}$ and denote it as 
$\partial \Sigma_{i}^{\rm vac}$.  
Topologically $\partial \Sigma_{i}^{\rm vac}=S^2$,
but {\em metrically}  $\partial \Sigma_{i}^{\rm vac}$ disappears in the
limit of shrinking the `thickened point set' down to $\{0\}\times S^{1}$:
the boundary theory on $\partial M_{i}^{\rm vac}$ is trivial, and the
corresponding `subtraction entropy' is thus zero.  On the other hand,
for the $d=4$ black hole, $\partial M_{i}=S^{1}\times S^{2}$,
$\partial\Sigma_{i}=S^2$, and ${\textstyle\rm Vol}\, (\partial\Sigma_{i})$
becomes 
the volume of the bifurcation sphere in
the limit of shrinking the thickened bifurcation sphere down to
$\{0\}\times S^{2}$, and this does {\it not} disappear.  
The boundary theory on $\partial M_{i}$ is {\it not}
trivial, and is responsible for the black hole entropy.  

I argued in Section~2 that this approach to black hole microstates is
fundamentally topological in nature, in that $\partial M_i$---the boundary
on which the physically relevant microstates live---arises from a certain 
sensitivity of $I^{(1)}[e]$ to the topology of $M$.
To clarify and strengthen this point recall
that the entropy predicted by the boundary theory on $\partial M_{i}$
is proportional to ${\rm Vol}\,(\partial\Sigma_{i})/\kappa$
(see (\ref{eqn:main_result})).
This is somewhat peculiar.  {\em Nowhere}
in the analysis leading to (\ref{eqn:main_result}) do we require a 
{\em metrical} property of the $S^1$ sector of 
$\partial M_{i}$ ($=S^{1}\times \partial\Sigma_{i}$), only a topological one:
the $S^1$ comes from
removing a single point ($\times S^{d-2}$); certainly there is a 
Euclidean time coordinate on this $S^{1}$, periodic with period equal to
the inverse temperature, but {\em metrically} $\partial M_{i}$ 
(but not $\partial\Sigma_{i}$) disappears in the limit.  So the physically
relevant microstates live on the boundary of 
a thickened bifurcation sphere, but
there is no length scale associated with this thickening.  It is not
`physics at the Planck scale' on a horizon interpreted as a `tangible'
boundary, but rather `physics at no scale' on a manifold whose 
{\it raison d'\^{e}tre} is topological in nature.

\vspace{1.5ex}
\begin{flushleft}
\large\bf Acknowledgments
\end{flushleft}
\noindent
I would like to thank Robert Mann, Gabor Kunstatter, and
Abhay Ashtekar for stimulating discussions, and each, including especially
Steven Carlip,
for reading the manuscript and providing 
insightful criticisms.  I also thank Jack Gegenberg for some discussions
during the early phase of this work.
This work was supported by grants from the Natural Sciences and
Engineering Research Council of Canada.

\end{document}